%% file: main.tex
\documentclass[conference]{IEEEtran}
\IEEEoverridecommandlockouts

\usepackage{listings}
\usepackage{chngcntr}
\usepackage{float}
\floatstyle{plain}
\newfloat{Code}{h!}{myc}
\newfloat{Table}{h!}{myc}
\newfloat{CodeSide}{}{myc}
\newfloat{CodeBottom}{b!}{myc}
\newfloat{CodeTop}{t!}{myc}
\usepackage{blindtext}
\usepackage[utf8]{inputenc}
\usepackage{comment}
\usepackage{xurl}

\floatstyle{plaintop}
\restylefloat{Table}
\usepackage{caption} 
\usepackage{cite}
\usepackage{amsmath,amssymb,amsfonts}
\usepackage{algorithmic}
\usepackage{graphicx}
\usepackage{textcomp}
\usepackage{xcolor}

\def\BibTeX{{\rm B\kern-.05em{\sc i\kern-.025em b}\kern-.08em
    T\kern-.1667em\lower.7ex\hbox{E}\kern-.125emX}}
\begin{document}
\title{Analysis of Validating and Verifying OpenACC Compilers 3.0 and Above\\

\thanks{\scriptsize
   This manuscript has been authored by UT-Battelle, LLC under Contract
   No.  DE-AC05-00OR22725 with the U.S. Department of Energy. The United States
   Government retains and the publisher, by accepting the article for
   publication, acknowledges that the United States Government retains a
   non-exclusive, paid-up, irrevocable, world-wide license to publish or
   reproduce the published form of this manuscript, or allow others to do so,
   for United States Government purposes. The Department of Energy will provide
   public access to these results of federally sponsored research in accordance
   with the DOE Public Access Plan
   (http://energy.gov/downloads/doe-public-access-plan).
   This paper is authored by an employee(s) of the United States Government and
   is in the public domain. Non-exclusive copying or redistribution is allowed,
   provided that the article citation is given and the authors and agency are
   clearly identified as its source.}
}

\author{\IEEEauthorblockN{1\textsuperscript{st} Aaron Jarmusch}
\IEEEauthorblockA{\textit{University of Delaware} \\
jarmusch@udel.edu}
\and
\IEEEauthorblockN{2\textsuperscript{nd} Aaron Liu}
\IEEEauthorblockA{\textit{University of Delaware} \\
olympus@udel.edu}
\and
\IEEEauthorblockN{3\textsuperscript{rd} Christian Munley}
\IEEEauthorblockA{\textit{University of Delaware} \\
chrismun@udel.edu}
\and
\IEEEauthorblockN{4\textsuperscript{th} Daniel Horta}
\IEEEauthorblockA{\textit{University of Delaware} \\
dchorta@udel.edu}
\and
\IEEEauthorblockN{5\textsuperscript{th} Vaidhyanathan Ravichandran}
\IEEEauthorblockA{\textit{University of Delaware} \\
vaidhy@udel.edu}
\and
\IEEEauthorblockN{5\textsuperscript{th} Joel Denny}
\IEEEauthorblockA{\textit{Oak Ridge National Laboratory} \\
dennyje@ornl.gov}
\and
\IEEEauthorblockN{6\textsuperscript{th} Sunita Chandrasekaran}
\IEEEauthorblockA{\textit{University of Delaware} \\
schandra@udel.edu}
\and
}

\maketitle
\thispagestyle{plain}
\pagestyle{plain}
\begin{abstract}
\input{Content/abstract}
\end{abstract}


\section{Introduction}
\input{Content/introduction}

\section{Overview of the Programming Model}
\input{Content/overview}

\section{Challenging Tests}
\input{Content/tests}

\section{Infrastructure}
\input{Content/infrastructure}

\section{LLVM Intergration}
\input{Content/llvm}

\section{Setup and Compilation Flags}
\input{Content/setup}

\section{Results}
\input{Content/results}

\section{Discussion}
\input{Content/discussion}

\section{Related Work}
\input{Content/related_work}

\section{Conclusion and Future Work}
\input{Content/conclusion}
\subsection{Acknowledgments}
\input{Content/ack}

\bibliographystyle{IEEEtran}
\vspace{12pt}
\bibliography{main}

\end{document}

%% file: Content/abstract.tex
OpenACC is a high-level directive-based parallel programming model that can manage the sophistication of heterogeneity in architectures and abstract it from the users. The portability of the model across CPUs and accelerators has gained the model a wide variety of users. This means it is also crucial to analyze the reliability of the compilers' implementations. To address this challenge, the OpenACC Validation and Verification team has proposed a validation testsuite to verify the OpenACC implementations across various compilers with an infrastructure for a more streamlined execution. This paper will cover the following aspects: (a) the new developments since the last publication on the testsuite, (b) outline the use of the infrastructure, (c) discuss tests that highlight our workflow process, (d) analyze the results from executing the testsuite on various systems, and (e) outline future developments.


%% file: Content/introduction.tex
Heterogeneous systems equipped with CPUs and accelerators are becoming increasingly prevalent. To add to the heterogeneity, application developers also have to face the challenges of migrating code from one type of heterogeneous system to another. Code migration is not only tedious but also prone to introducing bugs. While there are a variety of language and models available for the application developers to choose from, such as  OpenACC~\cite{OpenACCstandard}, OpenMP~\cite{OpenMP}, CUDA~\cite{cudasdk}, OpenCL~\cite{OpenCL}, NVIDIA Thrust~\cite{thrust}, and Kokkos~\cite{edwards2014kokkos} among there, it depends on what the developers are the most comfortable with and what would fetch them performance without losing portability, accuracy among other metrics of their applications' interest.  

In this paper, we focus on OpenACC, a directive-based programming model that targets traditional X86 architectures and accelerators such as GPUs and FPGAs~\cite{lee2016openacc}, validity of the OpenACC compiler implementations and their conformance to the standard specification; these will include features such as \texttt{atomic} or \texttt{parallel} constructs, clauses such as \texttt{if} or \texttt{copy\_out}, or combinations of occurrences of clauses on constructs. Support from Commercial compilers include NVIDIA and Hewlett Packard Enterprise (HPE Cray). Currently, NVIDIA supports up to version 2.7. HPE Cray focuses on Fortran and fully supports OpenACC 2.0 with partial support up to 2.6. These implementations are being developed for not only NVIDIA GPUs but also AMD GPUs as we are beginning to see a number of systems supporting the AMD GPUs including Frontier, the first exascale system. Open Source compilers include GNU Compiler Collection (GCC)\cite{openaccgcc} with support for OpenACC 2.6 and Clacc (OpenACC support for Clang and LLVM)~\cite{denny2018clacc}.  

Academic compilers\footnote{More information on existing compilers can be found on the OpenACC webpage https://www.openacc.org/tools} include Omni Compiler from Riken/University of Tsukuba~\cite{tabuchi2013source}, OpenARC from ORNL~\cite{lee2014openarc} and OpenUH from SBU and UH (that are now outdated)~\cite{wolfe2017implementing}.

OpenACC has been adopted for large-scale scientific applications on large-scale computing systems such as Summit at the Oak Ridge Leadership Facility and JUWELS at the Jülich Supercomputing Centre, among several others. Some of these applications include ANSYS1~\cite{sathe2016accelerating}, GAUSSIAN~\cite{gaussian}, ICON~\cite{sawyer2014towards}, COSMO 3~\cite{lapillonne2014using}, MPAS microphysics WSM6 ~\cite{kim2021gpu} and many more. Please refer to this tracker where OpenACC has been collecting published OpenACC papers~\footnote{shorturl.at/mquv0}.

To ensure application developers can successfully use the model for their large scale applications, it is critical that the implementations are validated and verified. The suite will not only check for conformance of the implementations to the specification but also push for consistency in functionality across implementations, which is key for the developers' success stories. It is a real challenge to debug if different implementations lead to different interpretations of the specification. This can only lead to more ambiguities among both the developers and the users. By creating functionality tests and amassing the results from multiple compilers/versions to compare implementations' adherence to the OpenACC specifications, this testsuite will enable the compiler developers to improve the quality of their tools and ensure compliance of their implementations with the specifications of the language. Our previous publication on this effort~\cite{wang2014validation} captured these discrepancies up to specification 2.5 (although at the time implementations did not fully cover 2.5 just yet). 

The specifications have since advanced to 3.2 with major updates including features such as reduction on arrays/composites, C++ lambda support, updated base languages, \texttt{acc\_memcpy\_d2d}, support for Fortran Do Concurrent and Block constructs, \texttt{async\_wait} on data regions, \texttt{acc\_wait\_any} and also providing support for both shared and discrete memory machines. So most, if not all, tests that this paper discusses have been written to adhere to the current specification. The paper also presents the infrastructure of the testsuite and validates most up-to-date versions of all available compilers including Clacc and HPE Cray's newer OpenACC implementations (supporting Fortran language).

The license associated with our testsuite is a dual license scheme. We are open to contributions from the community. The dual license scheme is designed in way to preserve the license used by the contributor, while the other license will be from OpenACC to ensure consistency in code versions, running code, and reporting of results. 
For more information on the project, how to contribute and our license please consult our website~\footnote{https://crpl.cis.udel.edu/oaccvv/}. 
The paper makes the following contributions:
\begin {itemize}
\item Provide a novel testing infrastructure for C/C++ and Fortran tests
\item Identify and report results on compiler bugs and runtime errors on all available and in use OpenACC compilers 
\item Evaluate different compilers' implementations for its conformance to the OpenACC specifications
\end{itemize}

Please refer to the project GitHub~\footnote{https://github.com/OpenACCUserGroup/OpenACCV-V} for all the codes in our testsuite.

%% file: Content/overview.tex
OpenACC is a programming model that can express high-level parallelization in three levels: gang, worker, and vector. The gang level is the broadest and comprises one or more workers, can be conceptualized as a thread block, and multiple gangs can work independently of one another. A worker is the middle level of parallelism or a block of vectors within a gang; the purpose of a worker is to compute a vector. Vector is the lowest level of parallelism, executing at the thread level, working in lockstep with one another.
OpenACC is designed to be a portable incremental optimization language, meaning the same code works on various architectures, parallelization can be added incrementally without changing the source code, and it is focused on providing comparable performance to lower-level paradigms. This is acomplizhed by abstracting the underlying hardware architecture away from the user allows for portability across several systems requiring minimal to no changes in the implementation. Focusing on the use of general-purpose graphics card units (GPUs), the model has, especially in recent years, broadened the number of systems that can utilize this programming model. 

To begin parallelizing, the user must tailor directive implementation based on the source code language. For C and C++, directives are prefixed by \texttt{\#pragma acc}. For Fortran, they are prefixed with the sentinel \texttt{!\$acc}. After the prefix, the user declares the implementation through various data and compute constructs using directives and clauses.

With proper implementation of OpenACC within a source code, the compiler handles parallelization and memory management. At compile time, the hardware architecture can be specified or detected by the compiler, and an architecture-specific optimized translation of the code is generated. At runtime, data is transferred to the device from the host as specified or implied by directives created in the source code, and the address is stored along with the host address to prevent a page error from the operating system.

\begin{Code}
\begin{lstlisting}[frame=single, caption=Simple Serial Addition of Arrays, label=Serial, numbers=none]
int main(){
    int N = 1<<20;
    ***Variable Declaration***
    for (int x = 0; x < N; ++x){a[x]=10;b[x]=15}
    for (int x = 0; x < 1 << 14; ++x){
        for (int y = 0; y < N; ++y){
            c[y] = a[y] + b[y];}}
    return 0;}
\end{lstlisting}
\end{Code}

Abstraction simplifies the creation and maintenance of optimized codes, and reduces the possibility of user error, by minimizing the time spent by the developer in implementing parallelization, and the steps required for implementation and optimizing on different architectures. As an example of the simplicity of translation of a serial source code into an OpenACC directive-based optimized code, we translate the serial code in Code \ref{Serial} to both CUDA and OpenACC.

\begin{Code}
\begin{lstlisting}[frame=single, caption=Simple CUDA Addition of Arrays, label=CUDA, numbers=none]
__global__
void add_arrays(int n, double  *a, double *b, double *c){
    int x = blockIdx.x * blockDim.x + threadIdx.x;
    if (x < n){c[x] = a[x] + b[x];}
}
int main(){
    int N = 1<<20;
    ***Variable Declaration***
    double *device_a, *device_b, *device_c;
    for (int x = 0; x < N; ++x){a[x] = 10;b[x] = 15;}
    cudaMalloc(&device_a, N * sizeof(double));
    cudaMalloc(&device_b, N * sizeof(double));
    cudaMalloc(&device_c, N * sizeof(double));
    cudaMemcpy(device_a, a, N * sizeof(double), cudaMemcpyHostToDevice);
    cudaMemcpy(device_b, b, N * sizeof(double), cudaMemcpyHostToDevice);
    for (int x = 0; x < 1 << 14; ++x){
        add_arrays<<<65565, 256>>>(N, device_a, device_b, device_c);
    }
    cudaMemcpy(c, device_c, N * sizeof(double), cudaMemcpyDeviceToHost);
    cudaFree(device_a);
    cudaFree(device_b);
    cudaFree(device_c);
}

\end{lstlisting}
\end{Code}

In order to translate this to CUDA, first, CUDA requires that we initialize pointers for the device data pointers. Each of these needs to be assigned its value by passing the reference to \texttt{cudaMalloc} which allocates the data on the device. After this, we also need to copy the data to the device referenced by the device pointers. To create the kernel, we define the function that accepts references and an end bound that checks against the \texttt{blockIdx.x}, \texttt{blockDim.x}, and \texttt{threadIdx.x}. The resulting code is in Code~\ref{CUDA}. Comparing the C code and the CUDA code, the two versions of the code require totally different function calls, formatting, and syntax.

\begin{Code}
\begin{lstlisting}[frame=single, caption=Simple OpenACC Addition of Arrays, label=ACC, numbers=none]
int main(){
    int N = 1<<20;
    ***Variable Declaration***
    for (int x = 0; x < N; ++x){a[x] = 10;b[x] = 15;}
    #pragma acc data copyin(a[0:N], b[0:N]) copyout(c[0:N])
        for (int x = 0; x < 1 << 14; ++x){
            #pragma acc parallel loop independent
            for (int y = 0; y < N; ++y)
                c[y] = a[y] + b[y];
}       }
\end{lstlisting}
\end{Code}

The OpenACC version in Code \ref{ACC} requires few changes to the serial code.  We add a \texttt{\#pragma acc data} construct, which manages the device data environment for that region, and a \texttt{\#pragma acc parallel loop independent} construct.  The \texttt{parallel loop} construct specifies that the following region should execute on the device and run in parallel, and the \texttt{independent} clause specifies that the loop does not have any inter-iteration dependencies.

%% file: Content/tests.tex
In order to validate and verify the conformance of compilers to the latest OpenACC specification, the testsuite must change over time. The OpenACC V\&V testsuite team is constantly updating the suite to test the most recent specification. In this section, we highlight some challenging tests within this update, specifically directives and clauses, that evoked varying interpretations.

\subsection{Routine Directive Bind Clause and Lambda Function}
\label{sec:routine}
The integration of additional C++ options with the \texttt{routine} directive and \texttt{bind} clause proved to be particularly tricky. These tests required several stages of development to guarantee tests exist for every combination of both C++ class objects with arrays and normal functions in tandem with lambda functions. 
The first obstacle is interpreting the specification. Reading over the relevant section, the \texttt{routine} directive was relatively straightforward. But for the \texttt{bind} clause, it is vague on both the implication and the use case. Thankfully, an imperfect test that utilizes these features exists but also creates a race condition. Utilizing the test to understand the directive application and implementation, the race condition was removed. This serves as a foundation for all of the other tests. Based on the specification, the \texttt{routine} directive allows the user to specify a function to be compiled and executed on both the host and on the device when the function is called on the host. The bind clause allows for further choice of which functions will be executed on the host versus the device.
The next challenge was learning how to correctly implement a C++ lambda function. So, the best next step that will provide practice without the extra complexity is testing its implicit behavior within a \texttt{pragma}. Interpreting it as an alternative way to declare functions that can exist within any scope, the integration was relatively straightforward. With the implementation of the routine and the lambda function understood, this stage of testing reveals several features that are not fully implemented. The current implementation of the lambda function requires the host function to not be a lambda function, the device lambda function to be placed at least one function declaration below where the prototyped \texttt{pragma} is declared, and the accelerator routine that is a lambda function must be defined after the \texttt{pragma} is declared. 

\begin{Code}
\begin{lstlisting}[frame=single, caption=Nonprototype Routine Declaration, label=nonprototype, numbers=none]
#pragma acc routine vector bind("device_array_array")
real_t host_array_array(real_t * a, long long n){
    #pragma acc loop reduction(+:returned)
    real_t returned = 0.0;
    for (int x = 0; x < n; ++x)
        returned += a[x];
    return returned;
}
auto device_array_array = [](real_t * a, long long n){
    real_t returned = 0.0;
    #pragma acc loop reduction(-:returned)
    for (int x = 0; x < n; ++x)
        returned -= a[x];
    return returned;
};
\end{lstlisting}
\end{Code}

\begin{Code}
\begin{lstlisting}[frame=single, caption=Prototype Routine Declaration, label=prototype, numbers=none]
real_t host_array_array(real_t *a, long long n);
#pragma acc routine(host_array_array) vector bind(device_array_array)

real_t host_array_array(real_t * a, long long n){
    #pragma acc loop reduction(+:returned)
    real_t returned = 0.0;
    for (int x = 0; x < n; ++x)
        returned += a[x];
    return returned;
}
auto device_array_array = [](real_t * a, long long n){
    real_t returned = 0.0;
    #pragma acc loop reduction(-:returned)
    for (int x = 0; x < n; ++x)
        returned -= a[x];
    return returned;
};
\end{lstlisting}
\end{Code}

Taking the four possible permutations, seen in Code \ref{nonprototype} and \ref{prototype}, of lambda functions and a normal function, these evaluate to a total of sixteen different declarations to integrate C++ class objects. Already having experience in using class objects, the last stage of development was less challenging. The implementation revealed another error, namely, utilizing a \texttt{copy} and \texttt{copyout} clause on a class object resulted in a segmentation fault when trying to access the data. But, this was worked around by utilizing the \texttt{update} directive with the \texttt{host} clause for the purpose of this test.

\subsection{If Clause}
\label{sec:if}
The \texttt{if} clause in OpenACC acts similarly to how if statements in the C language work. When the condition in the \texttt{if} clause evaluates to true, the region will execute, accelerated, on the device. When the condition evaluates to false, the region will run unaccelerated on the local host thread. Starting in the OpenACC 3.0 Specification, the \texttt{if} clause was applicable to the following directives: \texttt{init}, \texttt{set}, \texttt{shutdown}, and \texttt{wait}. The usage for the four directives is similar; the following test is for the \texttt{init} directive. When the \texttt{init} directive is called, the runtime environment for a given device is initialized. When we created the test for the \texttt{init} directive with an \texttt{if} clause, we originally believed the simplest way to test this was to create a compute region and see if it ran. We created two arrays and filled the first array, A, with random values, and the other array, B, with all zeros. We loop through array A and check if the value in A matched the same value in A, which would of course give us a true evaluation. Since the values would always match, we set the value in array B to the value in array A within an \texttt{init} \texttt{if} \texttt{pragma} that evaluated to true. We would then check if the values from both arrays always matched, and returned an error if they didn't.

Shortly after this code was created, through discussion an issue with our test was found; the \texttt{init} directive doesn’t deal with compute regions. One doesn't use \texttt{init} to compute any type of values, it is rather used to initialize the runtime for some device. Therefore, our original interpretation completely misused the directive. Connecting the \texttt{if} clause to an \texttt{init} directive only determines if runtime is initialized on either the current device or the local thread. Once we understood this, the test became much more clear, reference Code \ref{if}.

\begin{Code}
\begin{lstlisting}[frame=single, caption=Correct Usage of Init If, label=if, numbers=none]
	int device_num = acc_get_device_num(acc_get_device_type());
	#pragma acc init if(device_num == device_num)
\end{lstlisting}
\end{Code}

So, instead of computing values, we had to call a device type and, using the \texttt{if} clause, check if the device type we are calling matches the device type we are currently using. This is why there are two tests in this file: we had to check for a situation where the device types match and one where the device types do not match. The \texttt{if} clause follows a similar protocol for \texttt{set}, \texttt{shutdown}, and \texttt{wait}; a device will perform the action for the respective directive only if the current device type called matches the device type currently being used. The \texttt{if} clause is currently only implemented on the \texttt{wait} directive, but will be officially implemented on \texttt{init}, \texttt{set}, and \texttt{shutdown} in the near future.

\subsection{Reference Counter Zero Behavior}
\label{sec:copyout}
\begin{Code}
\begin{lstlisting}[frame=single, caption=Reference Counter Zero, label=copyout, numbers=none]
#pragma acc data copyin(a[0:n], b[0:n]) copy(c[0:n])
    #pragma acc parallel loop
    for (int x = 0; x < n; ++x)
        c[x] = a[x] + b[x];
#pragma acc exit data copyout(c[0:n])
for (int x = 0; x < n; ++x){
    if (fabs(c[x] - (a[x] + b[x])) > PRECISION)
        err += 1;
}
\end{lstlisting}
\end{Code}
The goal of this test, Code 7, is to verify that an \texttt{exit data} clause or \texttt{copyout} clause does not cause a runtime error when executed on a variable whose reference count is zero. In OpenACC 3.0 and earlier versions, this action should cause a runtime error, however, in OpenACC 3.1 and on the error should not occur and no action should be taken. To test this feature, a data region is created, and a variable, the c array, is copied in and then out of the region. On entrance to the data region, the reference count of the variable is incremented to one, and on exit, it is decremented to zero. After the data region, an \texttt{exit data copyout} directive is executed on the aforementioned variable whose reference count is now zero. This should cause no action, according to the update to the specification in OpenACC 3.1, whereas it would cause a runtime error in OpenACC 3.0 and before. This test was difficult because the reference counter is not directly accessible to the user of OpenACC directly; the programmer must trust the implementation of the compiler. After multiple revisions of this test, we decided the best way to test it was to use an \texttt{exit data} directive with a \texttt{copyout} clause after a data region, seen in Code \ref{copyout}. The redundant \texttt{copyout} clauses on the data region would not cause the reference count to go below zero. Moreover, clauses on a data region are not read by the compiler in the order that they are listed, so putting a \texttt{copyout} on a data region where the data to be copied is not already present can cause issues. An \texttt{exit data} clause following a region where the data has already been copied back to the host implies that the reference counter is zero on the variable which is attempted to copy again, thus testing the action taken when a \texttt{copyout} is performed on a variable whose reference count is zero.

\subsection{Init Directive Device Type and Num Clauses}
Initially looking into the \texttt{device type} clause, the feature was relatively straightforward to utilize with the function \texttt{acc\_get\_device\_type}. Making the assumption that the return value for the function could be used for the clause, we looked into the specification for further guidance and found that it accepts device-type-list. Looking throughout the specification, this value was never explicitly defined.

\begin{Code}
\begin{lstlisting}[frame=single, caption=device\_type for GCC, label=typeGCC, numbers=none]
int test1(){
    int err = 0;
    srand(SEED);
    int device_num = acc_get_device_num(acc_get_device_type());
    #pragma acc init device_num(device_num)
    return err;}
\end{lstlisting}
\end{Code}

But when using this implementation, Code \ref{typeGCC}, for an NVIDIA GPU, the compiler threw an error about \texttt{device type} requiring the keywords: host, multicore, default, or NVIDIA. To test this further, this code was also compiled for AMD GPUs using GCC resulting in no compilation or runtime errors. Rewriting the tests to incorporate the outlined changes from the compiler from NVIDIA, the new test, Code \ref{devicetype}, was compiled using nvc and GCC.

\begin{Code}
\begin{lstlisting}[frame=single, caption=device\_type for nvc, label=devicetype, numbers=none]
 int test1(){
    int err = 0;
    srand(SEED);
    #pragma acc init device_type(host)
    #pragma acc init device_type(multicore)
    #pragma acc init device_type(default)
    #pragma acc init device_type(nvidia)
    return err;}
\end{lstlisting}
\end{Code}

Asking the OpenACC community for further guidance led to the appendix section. This section outlines the different keywords to use for the different GPU vendors. While very useful, it does not fully encapsulate all of the possible keyword options for the GPUs with no mention of the multicore option for NVIDIA or the \texttt{acc\_get\_device\_type} option for GCC. This also resulted in the encapsulation of each of the options into their own tests and the NVIDIA option being isolated into a separate test file.

%% file: Content/infrastructure.tex
\begin{figure}
\caption{Overview of the infrastructure}
\label{html_output}
\includegraphics[width=\columnwidth]{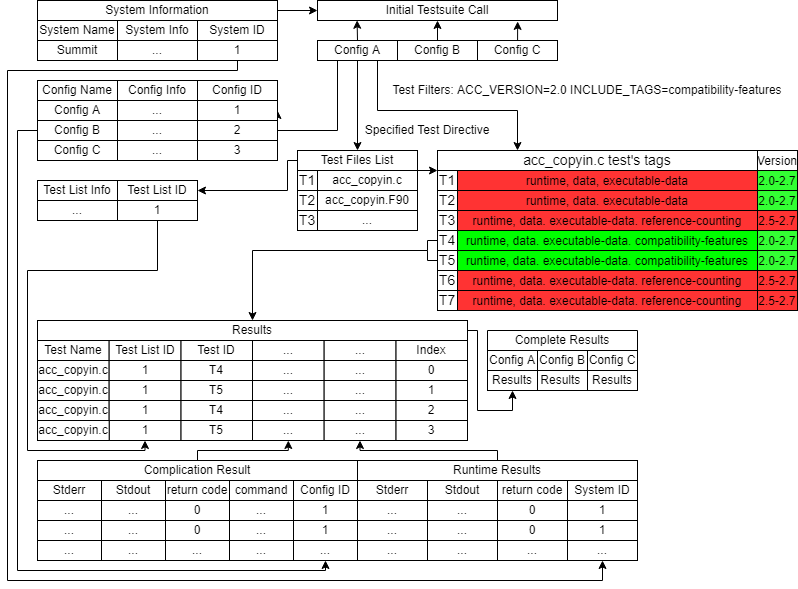}
\end{figure}

\label{sec:infrastructure}
Alongside the validation suite for OpenACC, an infrastructure has been created to provide tools to maximize customizability and readability. Formatting the output and results, it was designed to handle errors at compilation and/or at runtime, simplify running the testsuite, and set up the runtime environment. Written in python using packages supported by both python 2 and python 3, it has been developed for compatibility across a majority of systems. Figure~\ref{html_output} gives an overview of the infrastructure. 

To customize the infrastructure, the users may edit the configuration file. Within the file, documentation is provided for each feature. Some of the features are left intentionally empty to allow for the full capability of the infrastructure. The most important feature is the interchangeable compiler that is being used. For C, C++, and Fortran tests, the compiler must be specified individually. Compilation flags must also be specified as well, such as \textit{-acc}, which is required for OpenACC libraries. The users also have the option to choose among the flags to use on the command line. These include but are not limited to: OpenACC-specific flags like \textit{-Minfo=all}, displaying error messages during runtime or compilation, and output of results for each test file. It is also possible to specify commands to be run pre- or post compilation or execution if the user would like to further customize the environment or output format.

Next, the users can partition the testsuite to only execute a certain subsection of tests. This could be accomplished in one of two ways: running only a specified directory or using the tag-based system for conditional compilation. The directory method utilizes the organization of the testsuite to isolate a specific version of OpenACC tests to execute. For the tag-based system, each test is tagged at the beginning of the code based on what feature of OpenACC it is testing. It can also be used to isolate a version of OpenACC that the tests are valid in. If tests are tagged with a later version, they will be omitted. For both methods, the directory where the tests are built can be specified. When no specified directory or tag is provided, the infrastructure compiles and executes all of the tests within the validation suite. 

Another important feature of the configuration file is the ability to customize how the output is formatted by utilizing one of the current options: json, txt, or html. With the json option, the data is exported in json format with nothing omitted. With the txt option, a list of commands and results are output within a text log. The html option will produce an altered json format for the purpose of being used in an html page. This altered json format can be read with the use of the results template.

Lastly, the testsuite could take several hours to fully execute all of the tests provided. To cut down on the execution time, a timeout parameter feature is provided to prevent the suite from hanging in execution when a test hangs. Each test should complete within 5 seconds, so a timeout of 10 seconds is recommended.  However, the largest time complexity is n$^3$ for any test currently, so the timeout can be scaled proportionally if the problem size is changed by the user. 

\begin{Table}
\begin{center}
\setlength{\tabcolsep}{2pt} 
\begin{tabular}{| c | c|}

\hline
Compilation Flags & description \\
\hline
-c & specify one or more configuration files\\
\hline
-env & environmental output\\
\hline 
-o & naming the output file\\
\hline 
-verbose & interactive intuitive all\\
\hline 
-system & label system being used\\
\hline 

\end{tabular}
\caption{Compilation Flags}
\label{tab:runset}
\end{center}
\end{Table}
Table 1 describes some of the flags available to users. A comprehensive list of all the commands and how to run the testsuite can be found on the OpenACC V\&V Testsuite GitHub~\footnote{https://github.com/OpenACCUserGroup/OpenACCV-V}.

%% file: Content/llvm.tex
The primary goal of the integration is to facilitate the use of the OpenACC V\&V Suite for testing the behavior of Clacc's current and future OpenACC support~\cite{llvmintegration}.

We anticipate this testsuite project will immediately benefit the development of Clacc (OpenACC support for C/C++ in Clang) and Flacc (OpenACC support for Fortran in Flang). These efforts would also be beneficial to establish feedback from Clacc and Flacc to the OpenACC V\&V testsuite team to advance testing for the latest OpenACC specification versions implemented in LLVM.
Clacc and Flacc are developing their own test suites that is to be contributed to upstream LLVM. While the OpenACC V\&V Suite can be used for validating all OpenACC implementations, the Clacc and Flacc testsuites are not meant to be that general. 
For example, Clacc's test suites exercise Clacc-specific command-line options and source-to-source capabilities, which are beyond the scope of the OpenACC specification. The major value of the OpenACC V\&V Suite would be an objective, third-party assessment of LLVM's (Clacc) conformance to the OpenACC specification that can be compared with other OpenACC implementations.
LLVM has a testing infrastructure that contains regression tests and whole programs~\cite{llvmlit}. 
The regression tests (small tests assessing features of LLVM) are expected to always pass and should be run before every commit. The whole program’s tests are driven by the LLVM Integrated Tester tool, which is part of LLVM. 
\subsection{Implementation}
To set up an environment for the OpenACC validation and verification suite entail a set of requirements that justify the different features that we have implemented so far. These requirements are as follows:
\begin{enumerate}
\item As mentioned in~\cite{coti2020openacc} the Clacc compiler translates OpenACC to OpenMP in order to build upon the OpenMP support being developed for Clang and LLVM. To maximize reuse of the OpenMP implementation, Clacc performs this translation early, on the abstract syntax tree, which is the compiler front end’s internal representation of the source code. This translation is effectively a lowering of the representation and thus follows the traditional ordering of compiler phases. 
\item A CMakeList file has been created and included in this project. It is a generator of build systems and used as an entry point to our testsuite. It generates the Makefile which allows users to compile, run, and report test results. A set of make rules has been created for each purpose, together with a set of options that modify each rule’s behavior.
\item Those who would like to use this testsuite must be able to obtain and export compilation and results. Hence, the designed infrastructure should allow them to obtain results in either a json format or format for exporting to other analysis tools and scripts.
\end{enumerate}

Clacc is a project to develop OpenACC compiler and runtime support for C and C++ in Clang and LLVM, Clacc’s source code is available publicly on GitHub as part of the LLVM DOE Fork maintained by ORNL. Clacc should be built in the same manner as upstream LLVM when Clang and OpenMP support are desired~\cite{coti2020openacc}. This integration includes a CMakeLists for the OpenACC V\&V suite so it can be built as part of the LLVM test-suite. The CMakeLists will search for all C and C++ source files of the OpenACC V\&V suite, and compile and run them. Running llvm-lit (or "make check") will require a compatible accelerator on the running machine. LLVM OpenACC V\&V  Integration can be found~\footnote{ https://github.com/llvm/llvm-test-suite/tree/main/External}. With these requirements defined, we present our infrastructure in the rest of this section.

%% file: Content/setup.tex
\label{sec:setup}
\begin{Table}
\begin{center}
\setlength{\tabcolsep}{5pt} 
\begin{tabular}{|p{0.16\linewidth} | p{0.25\linewidth} | p{0.23\linewidth} | p{0.21\linewidth}|}
\hline
System & Hardware & Compiler & Flags \\
\hline 
DARWIN & NVIDIA T4 & nvc 21.9 & -acc=gpu\\
\hline
DARWIN & NVIDIA T4 & nvc 22.5 & -acc=gpu\\
\hline
DARWIN & NVIDIA Tesla T4 & GCC 10.1.0 & -fopenacc\\
\hline
DARWIN & NVIDIA Tesla T4 & GCC 11.3.0 & -fopenacc\\
\hline
DARWIN & NVIDIA Tesla T4 & GCC 12.1.0 & -fopenacc\\
\hline
Summit & NVIDIA Tesla V100 & nvc 20.9 & -acc=gpu\\
\hline
Summit & NVIDIA Tesla V100 & nvc 22.5 & -acc=gpu\\
\hline
Summit & NVIDIA Tesla V100 & GCC 10.2.0 & -fopenacc\\
\hline
Summit & NVIDIA Tesla V100 & GCC 12.1.0 & -fopenacc\\
\hline

Perlmutter & NVIDIA A100 Tensor Core & nvc 21.11 & -acc=gpu\\
\hline
Perlmutter & NVIDIA A100 Tensor Core & nvc 22.5 & -acc=gpu\\
\hline
Perlmutter & NVIDIA A100 Tensor Core & GCC 10.3.0 & -fopenacc\\
\hline
Perlmutter & NVIDIA A100 Tensor Core & GCC 11.2.0 & -fopenacc\\
\hline
Perlmutter & NVIDIA A100 Tensor Core & Clacc & -fopenacc\\
\hline

Spock & AMD MI100 & GCC 10.3.0 & -fopenacc\\
\hline
Spock & AMD MI100 & GCC 11.2.0 & -fopenacc\\
\hline
Spock & AMD MI100 & HPE 12.0.0 & -hacc,noomp\\
\hline
Spock & AMD MI100 & HPE 13.0.0 & -hacc,noomp\\
\hline

\end{tabular}
\caption{Compiler Versions and Platforms}
\label{tab:runset}
\end{center}
\end{Table}

Table 2 displays the various systems and compilers we used, as well as the hardware associated with each system and what flags were used to specify OpenACC. You can find more information on the system on our website. For more information on command-line options for running OpenACC applications and installation of Clacc, please go to the Clacc Github repository of the llvm-project.\footnote{https://github.com/llvm-doe-org/llvm-project/tree/clacc/main}

%% file: Content/results.tex
(Note: The compiler and their versions, the number of tests, and the OpenACC specification version used to generate results for this paper are as current as Aug 2022 (at the time of submitting this paper). Subjected to acceptance, we plan to update results with any updated compiler versions we would have access to at the time of submitting the final version of the manuscript.)

While hundreds of tests have been added to or modified in the testsuite, nothing has changed with the portability and ease of use. After adjusting to each system the suite has run with multiple compilers on multiple systems, when compatible, as described in the previous section: Setup and Compilation Flags in Section~\ref{sec:setup}. Over the past couple of years, the testsuite has grown to a suite comprised of more than 800 tests. Running these tests across multiple systems gave varying results. Out of the 830 tests, approximately 80\% passed with the NVIDIA and GNU compilers. Due to the renewed HPE Cray OpenACC compiler now focusing on the Fortran language, the success rate is higher at roughly 84\%. 

The success rate of each compiler will continue to evolve as more features are implemented per compiler and more bugs are fixed. We hope that the testsuite not only provides a metric for compiler teams to measure their current implementation against but also will be a tool in the future for compilers to verify that features continue to work as changes continue to be implemented. 

\begin{figure}
\caption{Verification and Validation Testsuite Results}
\label{Results_graph}
\includegraphics[width=\columnwidth]{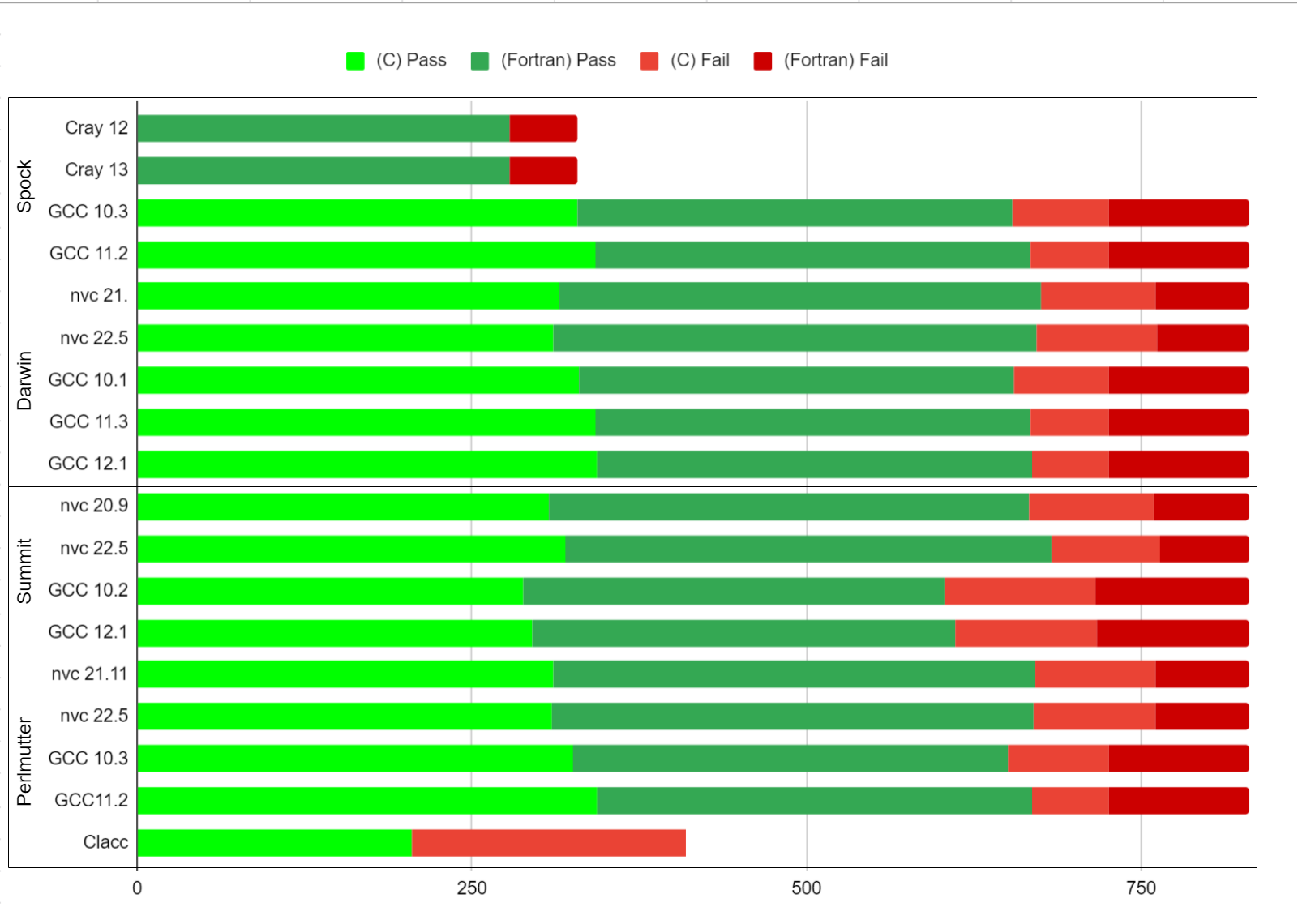}
\end{figure}

\subsection{NVIDIA Compiler's Conformance to the OpenACC Specifications}
The NVIDIA compiler was tested on three different systems intentionally to identify variations in results. Utilizing the latest available compiler version on each system, namely 22.5, and an older version to demonstrate validity over time, no significant variation between versions tested was seen. On average in these results, the NVIDIA compiler supports 80.7\% of features currently implemented into the testsuite which includes most features up to OpenACC 3.2. 

Some of the failures are due to issues in the implementations of OpenACC; issues have been reported to NVIDIA who have logged bugs. These include but are not limited to: the~\texttt{init} directive with the~\texttt{device\_type} clause, a combined construct for~\texttt{serial loop} with the~\texttt{vector},~\texttt{worker}, and~\texttt{gang} clauses. Other failures are because some features past OpenACC 2.7 are not yet implemented. The~\texttt{if} clause on the \texttt{init},~\texttt{set}, and~\texttt{shutdown} directives to name a few. NVIDIA is aware of all failures and are continually working to improve the compiler. 

\subsection{The GNU Compiler's Conformance to the OpenACC Specifications}
The GNU compilers support anywhere between 78.3\% and 80.3\% of the features currently implemented. The only outlier was on Summit, which supported between 72.6\% and 73.5\% of features currently implemented with GNU compilers. These features in the testsuite include most of the features implemented up to OpenACC 3.2 and were written in C and Fortran. While the number of tests written in C that passed varied per machine, the results were consistent with each other. The Fortran tests had almost the exact same pass rate, with again Summit being the only exception for files in both languages. Most of these failures were due to either bugs in the compiler or the compiler not fully implementing all features up to OpenACC 3.2. 

\subsection{The HPE Cray Fortran Compiler's Conformance to the OpenACC Specifications}

One of the recent addition to the suite is HPE Cray's OpenACC support for Fortran. To note, this compiler focuses on the Fortran language, thus limiting the tests we were able to run. Our infrastructure will exclude the other two languages when running, the option is within the config file. Once excluded, the testsuite is left with a total number of 329 tests in Fortran with a success rate of approximately 84\%. To achieve this success rate we used the pre-exascale system called Spock~\cite{spock} at Oak Ridge National Laboratory. This system has nodes with AMD GPUs MI60 and MI100. Our tests ran on the  MI100. 

The HPE Cray compiler uses PTX (Parallel Thread Execution) instructions which are translated into assembly, it does not translate into CUDA. However, as clarified in previous sections the Cray compiler fully supports OpenACC 2.0, with partial support for OpenACC 2.6. This includes many behaviors like \texttt{copy} data clauses and the \texttt{default} clause. Our testsuite has confirmed support for the \texttt{atomic} construct, the \texttt{reduction} clause with the \texttt{kernels} and \texttt{parallel} constructs. Notably, since partial support is up to 2.6 some test cases will fail. Some of these cases include: \texttt{acc\_on\_device}, \texttt{acc\_set\_device\_num}, and  \texttt{acc\_set\_device\_type}. We cannot guarantee that these tests will succeed in the next compiler update. We can guarantee that our team will continue to look into detail and communicate with the developer about these results. 

\subsection{The LLVM/Clacc Compiler's Conformance to the OpenACC Specifications}
The Clacc compiler focuses on the C and C++ language, thus limiting the tests we were to run. Our infrastructure will exclude the Fortran language when running, the option is within the config file. It was tested on NERSC’s Perlmutter using the LLVM GitHub commit hash "4879e96", which supported between 50\% of features currently implemented with Clacc compilers. These features in the testsuite include most of the features implemented up to OpenACC 3.2 and were written in C. Most of these failures were due OpenACC functions yet to be implemented into the compiler, these include: \texttt{kernels}, \texttt{serial}, \texttt{async/wait}. The Clacc compiler has greatly increased their support for OpenACC lately, however this is work in progress. The paper subjected to acceptance, we might be able to update the Clacc results for the final version of the paper as we iterate through their implementations.

%% file: Content/discussion.tex
This project has demonstrated the conformance of different versions of OpenACC compilers and their current performance targeting different systems including OLCF's Summit and Spock, NERSC's Perlmutter, and the University of Delaware's Darwin. The previously published version of this effort~\cite{10.1007/978-3-319-67630-2_39} had deemed the HPE Cray and Clacc compilers were not ready for validation. However, the two compilers are now showing increased support for OpenACC. 

We were able to target multiple systems due to the infrastructure that we have built since our last publication. Because the number of tests within our suite has increased and many users may want to run the full suite, the infrastructure will be the easiest way to validate any compiler on any system. One thing to note, we want to update our implementation of the infrastructure to better run C++ tests as currently, the infrastructure leaves these tests out of the results.

Recently, features newly added in versions beyond 3.0 may not have compiler implementations as of yet. However, we have written the tests  awaiting implementation. 
For example, the \texttt{if} clause was recently added to the \texttt{init}, \texttt{shutdown}, \texttt{set}, and \texttt{{}wait} directives in OpenACC 3.0. At the moment of this publication, only the wait directive with the \texttt{if} clause has been implemented by the NVIDIA (nvc) compiler. \texttt{Init}, \texttt{shutdown}, and \texttt{set} have yet to be implemented but the testsuite contains a test for each of these three combinations. This is the major reason why new tests are failing; however, as compilers will develop they will start to implement these newer features. Thus, our test will validate the compiler implementation of those newly compiler-implemented features. 


%% file: Content/related_work.tex
In our past publications on this topic, we have written about the testsuite covering OpenACC 1.0~\cite{wang2014validation} and OpenACC 2.5~\cite{10.1007/978-3-319-67630-2_39}. These two publications are the most related to our ongoing work. There is testsuite similar to ours, as at the University of Delaware, which targets the OpenMP programming model. Their validation and verification testsuite~\cite{OpenMPVV} also aims to check for conformance of features in compiler implementations. Since 2003, the OpenMP testsuite team has developed tests according to OpenMP 2.0~\cite{muller2003openmp}, version 2.5~\cite{muller2004validating}, version 3.1~\cite{wang2012openmp} and version 4.5~\cite{diaz2018openmp,diaz2019analysis}. The team works on the next publication as we speak. 

Other efforts towards creation of a validation and verification testsuite include Csmith~\cite{yang2011finding} and the Parallel Loops testsuite~\cite{dongarra1991parallel}, modeled after the Livermore Fortran kernels~\cite{mcmahon1986livermore} as also mentioned in our previous publication. Csmith uses differential testing to perform a randomized test-case generator exposing compiler bugs and the Parallel Loops testsuite chooses a set of routines to test the strength of a computer system (compiler, run-time system, and hardware) in a variety of disciplines. OvO~\cite{ovo} is another suite that is a collection of OpenMP offloading tests for C++ and FORTRAN. OvO is focused on testing extensively hierarchical parallelism and mathematical functions.

%% file: Content/conclusion.tex
The purpose of this project is to develop test cases in order to validate and verify compilers' implementations of OpenACC features. Compilers and the testsuite must advance alongside the evolving OpenACC specification. The evolution of the testsuite gives the opportunity to create test cases and corner cases resulting in the finding of bugs, thus, improving the compilers' implementation of each feature. 

In the future, one goal is to create an example guide for the OpenACC community. Taking inspiration from OpenMP 4.5 Examples, our hope is that the guide will display relatively simple examples of features in OpenACC. We will kick off this process by first surveying the OpenACC user community and seeking their input on features they would like us to create examples for. 
The goal of this document would be to explain each directive and clause with a short statement outlining what the code is accomplishing.  Automating the process gives the opportunity to develop the guide at a faster pace; however, our main focus, first, is to present a guide of the utmost quality to the community. Our main hope for the example guide is to provide sample code for OpenACC users to get a basic understanding of applying OpenACC to their codes.

In order to ensure that compiler implementations are as bug-free as possible and conform to the OpenACC specification, we will add more rigorous testing to cover edge cases that may not be possible with basic tests. To make the testsuite easy to use, we will create a transposable connection between the testsuite and the specification. The testsuite repository will have tests that are related to the specification, and each test will be tagged with the corresponding definition.

The results shown have been taken relatively close to the submission date of this paper. We will continue to develop the testsuite with valuable feedback and input from our mentors. For more details on up-to-date results with new compilers, compiler versions, up-to-date tests or to contribute to this project please consult our website\footnote{https://crpl.cis.udel.edu/oaccvv}. 

%% file: Content/ack.tex
We are grateful to OpenACC for their support on this project, including technical support from Mathew Colgrove, Jeff Larkin, Duncan Poole, Christophe Harle, Guray Ozen, Wael Elwasif, Seyong Lee and Joel Denny for continued help with this project. 

This research was supported by the National Science Foundation (NSF) under grant no. 1919839, in part through the use of DARWIN computing system: DARWIN – A Resource for Computational and Data-intensive Research at the University of Delaware. This material is also based upon work supported by NSF under grant no. 1814609.

This research used resources of the Oak Ridge Leadership Computing Facility at the Oak Ridge National Laboratory, which is supported by the Office of Science of the U.S. Department of Energy under Contract No. DE-AC05-00OR22725.

This research used resources of the National Energy Research Scientific Computing Center (NERSC), a U.S. Department of Energy Office of Science User Facility located at Lawrence Berkeley National Laboratory, operated under Contract No. DE-AC02-05CH11231 using NERSC award.